\documentstyle[twoside,fleqn,espcrc2,epsf]{article}

\newcommand{\AmS}{{\protect\the\textfont2
  A\kern-.1667em\lower.5ex\hbox{M}\kern-.125emS}}

\def\nn{\nonumber}
\def\msbar{\overline{\mbox{MS}}}
\def\gev{\mbox{GeV}}
\def\mev{\mbox{MeV}}

\title{Neutral $B$ Meson Mixing and Heavy-Light Decay Constants from
       Quenched Lattice QCD
\thanks{CPT-98/PE.3689; CERN-TH/98-289. 
Support from EPSRC and PPARC under grants GR/K41663 and GR/L29927
is acknowledged.}}

\author{Laurent Lellouch
        \address{Theory Division,CERN,CH-1211 Geneva 23,
        Switzerland}
        \thanks{On leave from: Centre de Physique Th\'eorique, Case
        907, CNRS Luminy, F-13288 Marseille Cedex 9, France.}%
        and 
        C.-J. David Lin\address{Department of Physics $\&$ Astronomy, 
        The University of Edinburgh, Edinburgh EH9 3JZ, Scotland}
        \thanks{Presenter.}
        (UKQCD Collaboration)}
       
\begin{document}

\begin{abstract}
We present high-statistics results for neutral $B$-meson mixing and 
heavy-light-meson leptonic decays
in the quenched approximation from tadpole-improved clover actions
at $\beta = 6.0$ and $\beta = 6.2$.  We consider quantities such as 
$B_{B_{d(s)}}$, $f_{D_{d(s)}}$, $f_{B_{d(s)}}$ and the full
$\Delta B=2$ matrix elements as well as the 
corresponding $SU(3)$-breaking ratios.  These quantities are important for
determining the CKM matrix element $|V_{td}|$.

\end{abstract}

\maketitle

\section{INTRODUCTION}

The study of $B^{0}_{d}-\bar{B}^{0}_{d}$ oscillations allows a clean 
extraction of the poorly known CKM matrix element $|V_{td}|$.  However,
the accuracy of this determination is currently limited by the theoretical
uncertainy in the calculation of the matrix element,
\begin{eqnarray}
{\mathcal{M}}_{bd} &=& \langle \bar{B}^{0}_{d} | 
 {\mathcal{O}}^{\Delta B = 2}_{d} | B^{0}_{d} \rangle \nn\\  
 &=& \langle \bar{B}^{0}_{d} | \bar{b} \gamma_{\rho} (1-\gamma_{5}) d
\bar{b} \gamma^{\rho} (1-\gamma_{5}) d | B^{0}_{d} \rangle \nn
\ ,\end{eqnarray}
which is related to the mass difference of the two mass eigenstates
of the $B^{0}_{d}-\bar{B}^{0}_{d}$ system, 
\begin{eqnarray}
\Delta m_d &=& \frac{G^{2}_{F}}{8\pi^2}M^{2}_{W}|V_{td}V_{tb}^{*}|^{2}
S_{0} \left(\frac{m^{2}_{t}}{M^{2}_{W}}\right)\eta_{B}
C_{B}(\mu) \nn\\ &\times& 
\frac{1}{2M_{B_{d}}}\left |{\mathcal{M}}_{bd}(\mu)\right | \nn
\ ,
\end{eqnarray}
where $G_{F}$ is the Fermi constant, $M_{W}$ the $W$-boson mass, $m_t$
the top-quark mass and $\mu$ the renormalisation scale. 
$S_{0}(m^{2}_{t}/M^{2}_{W})$, $\eta_{B}$ and $C_{B}(\mu)$ are 
perturbatively-calculated quantities.

An alternative approach, in which many theoretical uncertainties cancel,
is to look at the ratio
\[
%\begin{equation}
\frac{\Delta m_{s}}{\Delta m_{d}} = 
\left | \frac{V_{ts}}{V_{td}} \right |^{2} \frac{M_{B_{d}}}{M_{B_{s}}}
\left| \frac{{\mathcal{M}}_{bs}}{{\mathcal{M}}_{bd}}\right |
\equiv \left |\frac{V_{ts}}{V_{td}}\right |^{2}
\frac{M_{B_{d}}}{M_{B_{s}}} r_{sd} \,
\]
%\end{equation}
%
where $M_{B_{d(s)}}$ is the meson mass.  The matrix elements 
${\mathcal{M}}_{bd(s)}$ can be parameterised as 
%
%\begin{equation}
\[
{\mathcal{M}}_{bd(s)} = \frac{8}{3}f^{2}_{B_{d(s)}}M^{2}_{B_{d(s)}}
B_{B_{d(s)}}
\]
%\end{equation} 
%
where $f_{B_{d(s)}}$ is the decay constant, 
and $B_{B_{d(s)}}$ the $B$-parameter of $B^{0}_{d(s)}$ mesons.

In this work, we obtain the ratio $r_{sd}$ from the direct calculation
of ${\mathcal{M}}_{bs}/{\mathcal{M}}_{bd}$ as well as from the 
calculations of $f_{B_{s}}/f_{B_{d}}$ and 
$B_{B_{s}}/B_{B_{d}}$.

\section{SIMULATION DETAILS}

We use the tadpole-improved Sheikholeslami-Wohlert (SW) quark action,
%
%\begin{equation}
\[
S^{SW}_{F} = S^{W}_{F} - ig_{0}c_{SW}\frac{\kappa}{2}\sum_{x,\mu,\nu}
\bar{q}(x)P_{\mu \nu}(x)\sigma_{\mu \nu}q(x)
\]
%\end{equation}
%
to perform simulations on a $24^{3} \times 48$ lattice at 
$\beta = 6.2$ and a $16^{3} \times 48$ lattice at $\beta = 6.0$. 
Here $S^{W}_{F}$ is the standard Wilson action, $g_{0}$ the bare
gauge coupling, $c_{SW}$ the clover coefficient, $\kappa$ the
hopping parameter, and $P_{\mu \nu}$ a lattice definition of the 
gauge-field strength tensor.  Table \ref{tab:simparam} gives the simulation
parameters.  We use KLM normalisation for the quark fields.

\begin{table*}[hbt]
\setlength{\tabcolsep}{0.5cm}
%\newlength{\digitwidth} \settowidth{\digitwidth}{\rm 0}
%\catcode`?=\active \def?{\kern\digitwidth}
\caption{Simulation parameters. 
$\kappa_Q$ and $\kappa_q$ are the heavy- and 
light-quark hopping parameters.}
\label{tab:simparam}
\begin{tabular*}{\textwidth}{@{}l@{\extracolsep{\fill}}ccccc}
\hline 
\hline
$\beta$ & $\#$ configs. & $c_{SW}$ & $\kappa_{q}$ & $\kappa_{Q}$ \\
\hline
6.0 & 498 & 1.48 & 0.13700 0.13810 0.13856 & 0.114 0.118 0.122 0.126 0.130 \\
6.2 & 188 & 1.44 & 0.13640 0.13710 0.13745 & 0.120 0.123 0.126 0.129 0.132 \\
\hline \hline
\end{tabular*}
\end{table*}

\section{OPERATOR MATCHING}

Matching onto the $\msbar$ scheme 
is performed at one-loop in perturbation theory
using the coupling $\alpha_{\tiny{\msbar}}(\mu)$ defined from the
plaquette \cite{LM}.  
Since the  clover-leaf interaction term is proportional to 
$g_{0}$, 
we can use the perturbative results obtained from a tree-level clover 
action \cite{frezzotti} with modifications appropriate for
tadpole-improvement and KLM normalisation.

For the matching of four-fermion operators, we use the basis
\begin{eqnarray}
{\mathcal{O}}_{1}^{lat} &=& \gamma_{\mu} \times \gamma_{\mu} + 
    \gamma_{\mu}\gamma_{5} \times \gamma_{\mu}\gamma_{5},\nonumber\\
{\mathcal{O}}_{2}^{lat} &=& \gamma_{\mu} \times \gamma_{\mu} -
    \gamma_{\mu}\gamma_{5} \times \gamma_{\mu}\gamma_{5},\nonumber\\
{\mathcal{O}}_{3}^{lat} &=& I \times I + \gamma_{5} \times \gamma_{5}
   ,\nn\\
{\mathcal{O}}_{4}^{lat} &=& I \times I - \gamma_{5} \times \gamma_{5}
   ,\nonumber\\
{\mathcal{O}}_{5}^{lat} &=& \sigma_{\mu\nu} \times \sigma_{\mu\nu}
   .\nonumber
\label{eq:four_q_op}
\end{eqnarray}

We set the coupling and matching scales to $\mu = \frac{1}{a}$ and,
for consistency with the
literature, run divergent operators to 5 GeV, 
using 2-loop continuum RG in the $\msbar$
scheme with the appropriate number of flavours.

To estimate the systematic error associated with the one-loop
matching, we vary the scale $\mu$ in a range from $1/a$ to $\pi/a$.
Decay constants are not affected since they are normalized by $f_\pi$
and $B$-parameters change by about 3\% ($f_{\pi}$ varies by approximately
3\%). Since we are mainly interested
in $SU(3)$-breaking ratios for which these effects are even smaller, 
we neglect these small variations in what follows.

\section{ANALYSIS AND RESULTS}

We determine $\kappa_c$ and $\kappa_s$ from pseudoscalar meson masses.
We set the scale with $M_{\rho}$ for spectral quantities 
and $f_{\pi}$ for decay constants. In fact, 
these two quantities yield remarkably
similar scales.
(See Table \ref{tab:spectrum}.)  
\begin{table}[hbt]
%\setlength{\tabcolsep}{0.5cm}
%\newlength{\digitwidth} \settowidth{\digitwidth}{\rm 0}
%\catcode`?=\active \def?{\kern\digitwidth}
\caption{Critical and strange hopping parameters and inverse lattice
spacings.}
\label{tab:spectrum}
\begin{tabular}{ccc}
\hline
\hline
$\beta$ & 6.0 & 6.2 \\
\hline
$\kappa_{c}$ & 0.13924(1) & 0.13793(1) \\
$\kappa_{s}$ & 0.13757(8) & 0.13670(9) \\
$a^{-1}(M_{\rho}) (\gev)$ & 1.96(5) & 2.57(8) \\
$a^{-1}(f_{\pi}) (\gev)$ & 1.92(4) & 2.58(9) \\
\hline
\hline
\end{tabular}
\end{table}
We then linearly extrapolate and interpolate heavy-light decay constants,
$B$-parameters and $\Delta B=2$ matrix elements
to $\kappa_c$ and $\kappa_s$, keeping $\kappa_c$,
$\kappa_s$, $aM_{\rho}$ and $af_{\pi}$ in the bootstrap loop.
Fig.~1 shows examples of these extrapolations.
\begin{figure}[hbt]
\epsfxsize=8cm
\epsffile{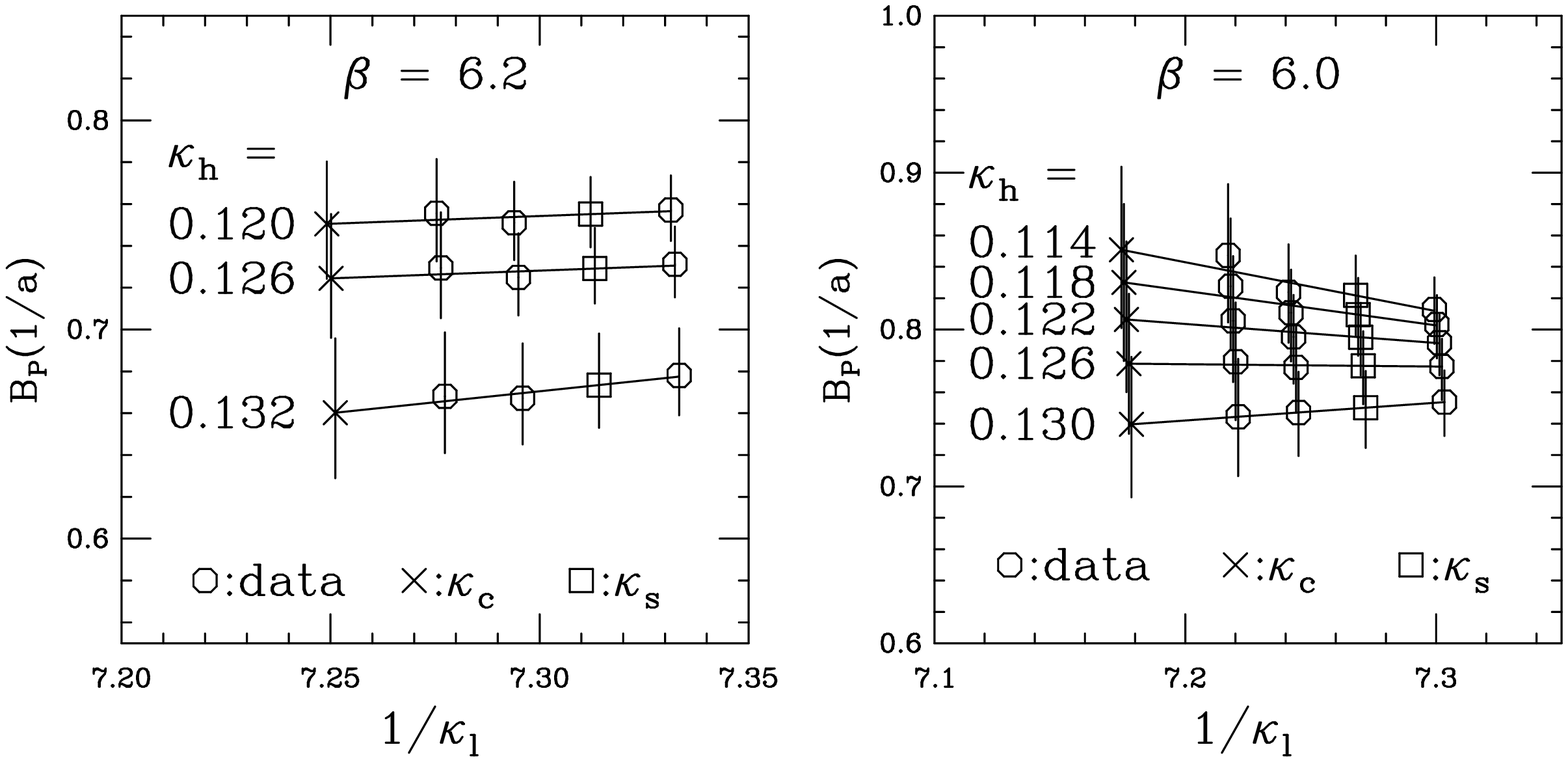}
\label{fig:Bvslqm}
\vspace{-24pt}
\caption{Light-quark-mass dependence of the heavy-light $B$-parameter, 
$B_P$, and extrapolation (interpolation) to $\kappa_l=\kappa_{c}$ 
($\kappa_s)$ at $\beta=6.0$ and 6.2.}
\end{figure}

For heavy-quark (HQ) extrapolations, we define ($M_P$ is the 
heavy-light meson mass)
\begin{eqnarray}
 \Phi_{f}(M_{P}) &=& \frac{af_{P}}{Z_{A}} \sqrt{\frac{M_{P}}{M_{\rho}}}
 \left \{ \frac{\alpha_{s}(M_{P})}{\alpha_{s}(M_{B})}
 \right \} ^{\frac{2}{11}}
  \nonumber \\ 
 \Phi_{\Delta F = 2}(M_{P}) &=&
  \frac{a^{4}{\mathcal{M}}M_{\rho}}{M_{P}}
 \left \{ \frac{\alpha_{s}(M_{P})}{\alpha_{s}(M_{B})}
 \right \} ^{\frac{4}{11}}
  \nonumber
\ .\end{eqnarray}
Then for $X(M_{P}){=}\Phi_{f}(M_{P})$,$\Phi_{\Delta F = 2}(M_{P})$, 
$B(M_{P})$ and $SU(3)$-breaking ratios, HQET predicts   
\[
 X(M_{P})= A_{X}\left \{ 1 + B_{X}(\frac{M_{\rho}}{M_{P}}) +
                  C_{X}(\frac{M_{\rho}}{M_{P}})^{2} + ...
 \right \}
. 
\]
Fig.~2 shows examples of the HQ extrapolations.
\begin{figure}[t]
%\vspace{9pt}
\epsfxsize=8cm
\epsffile{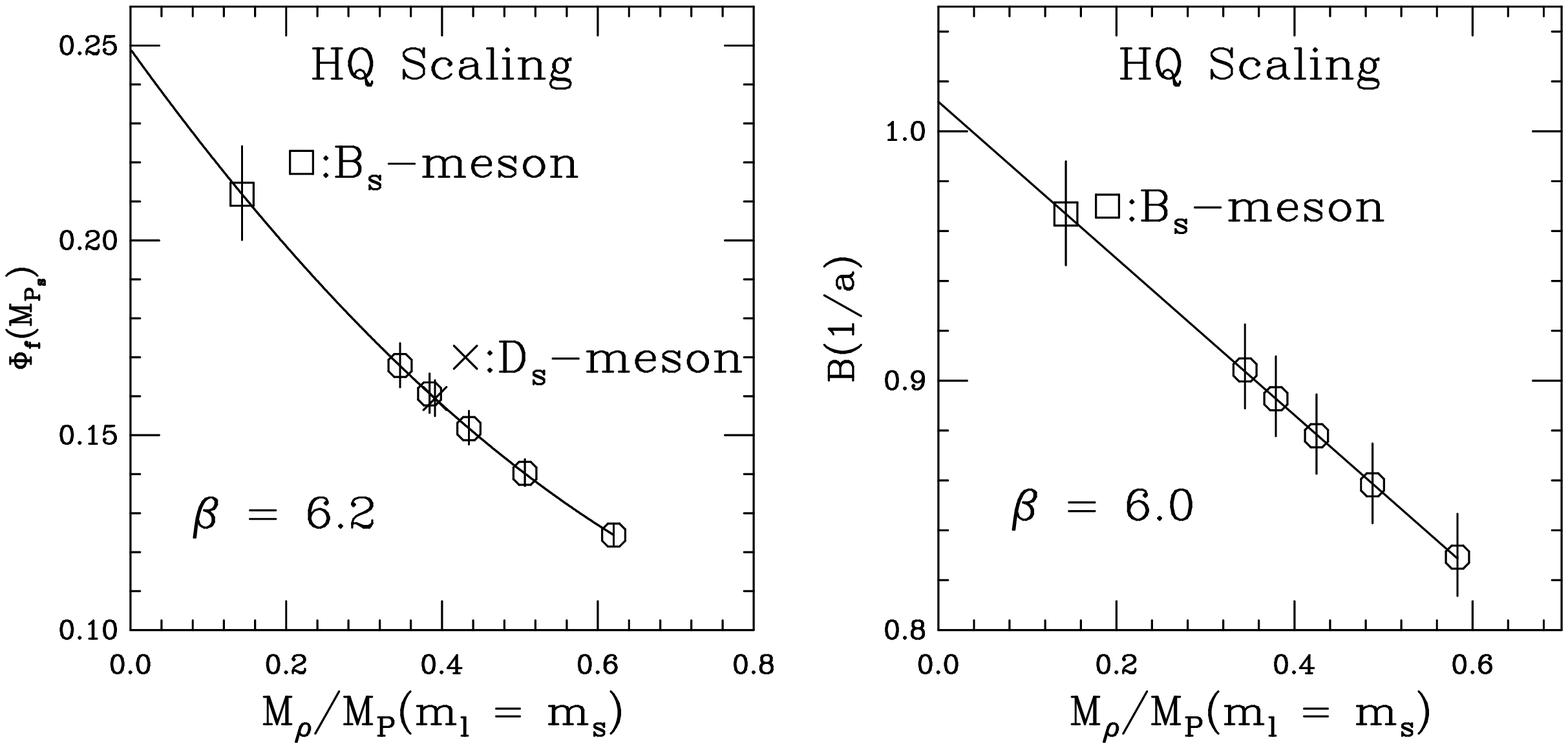}
\label{fig:PhiBHQ}
\vspace{-24pt}
\caption{HQ scaling of $\Phi_{f}$ and $B$-parameter.}
\end{figure}

For $SU(3)$-breaking ratios, we find that taking the ratio before or
after the HQ extrapolation leads to nearly indistinguishable
results. We use the former for our final results since
$SU(3)$-breaking ratios have milder HQ-mass dependences.

Our main results are summarised in Table \ref{tab:results}.  
We obtain $r_{sd}$ from the direct
calculation of ${\mathcal{M}}_{bs}/{\mathcal{M}}_{bd}$ as well
as from $f_{B_{s}}/f_{B_{d}}$ and
$B_{B_{s}}/B_{B_{d}}$.  Our results for the direct calculation
are consistent with those
of \cite{BBS}, obtained with propagating Wilson quarks, and, at $\beta=6.0$,
with the static result of \cite{guido}. 
However, as Fig.~3 suggests, it is more difficult to
control the chiral and HQ extrapolations of the matrix elements in the
direct calculation because these extrapolations are more pronounced.
\begin{table}[th]
\setlength{\tabcolsep}{0.5cm}
%\newlength{\digitwidth} \settowidth{\digitwidth}{\rm 0}
%\catcode`?=\active \def?{\kern\digitwidth}
\caption{Summary of results. Errors are statistical only.}
\label{tab:results}
\begin{tabular}{cccccc}
%{\textwidth}{@{}l@{\extracolsep{\fill}}rrrr}
\hline 
\hline
$\beta$ & 6.0 & 6.2 \\
\hline
$f_{D_{s}} (\mev)$ & 239(6) & 221(9) \\
$f_{D} (\mev)$ & 213(6) & 193(10) \\
$f_{B_{s}} (\mev)$ & 221(7) & 190(12) \\
$f_{B} (\mev)$ & 191(10) & 161(16) \\
$\frac{f_{D_{s}}}{f_{D}}$ & 1.12(1) & 1.15(4) \\
$\frac{f_{B_{s}}}{f_{B}}$ & 1.15(4) & 1.18(8) \\
$B_{B_{s}}^{\mathrm nlo} (5\gev)$ & 0.86(2) & 0.85(2) \\
$B_{B}^{\mathrm nlo} (5\gev)$ & 0.83(4) & 0.85(3) \\
$\frac{B_{B_{s}}}{B_{B}}$ & 1.03(3) & 0.99(3) \\
$(\frac{M_{B_{s}}}{M_{B}} \frac{f_{B_{s}}}{f_{B}})^{2}
 \frac{B_{B_{s}}}{B_{B}}$ & 1.38(7) & 1.37(13) \\
$\frac{{\mathcal{M}}_{bs}}{{\mathcal{M}}_{bd}}$ & 1.52(19) & 1.70(28) \\
\hline \hline
%\vspace{-12pt}
\end{tabular}
\end{table}
\begin{figure}[t]
%\vspace{9pt}
\epsfxsize=8.6cm
\epsffile{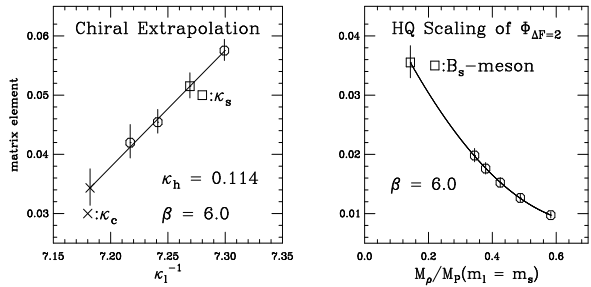}
\label{fig:ME_plot}
\vspace{-24pt}
\caption{Chiral and HQ extrapolation of the matrix element.}

\end{figure}

Because we have results at only two values of the lattice spacing, we
cannot extrapolate to the continuum limit. We therefore consider the
$\beta=6.2$ results to be our best, noting that decay constants may
still suffer from relatively large discretisation errors (roughly a $2\sigma$
effect between 6.0 and 6.2) while $SU(3)$-breaking ratios and $B$-parameters
are consistent within errors at the two $\beta$ values.

Although formally one need not include the $a\partial_{\mu} P$ correction to
the axial current when using a mean-field improved, tree-level clover action,
it would be interesting to investigate its effect on our results in view
of understanding how non-perturbatively, $O(a)$-improved decay constants
may behave. We plan to do so in the future.

For a comparison of our results with other recent results, we refer the reader
to \cite{terry}.

\end{document}